\documentclass[letterpaper,onecolumn]{IEEEtran}
\IEEEoverridecommandlockouts

\usepackage[utf8]{inputenc} 
\usepackage[T1]{fontenc}
\usepackage{url}
\usepackage{cite}
\usepackage[cmex10]{amsmath}
\usepackage{amssymb,amsfonts}
\usepackage{xfrac}
\usepackage{algorithmic}
\usepackage{graphicx}
\usepackage{textcomp}
\usepackage{xspace}
\usepackage{paralist}
\usepackage[dvipsnames]{xcolor}

\usepackage{hyperref}
\hypersetup{
  linkcolor  = Violet!85!black,
  citecolor  = YellowOrange!85!black,
  urlcolor   = Aquamarine!85!black,
  colorlinks = true,
}
\usepackage[capitalize,nameinlink]{cleveref}
\crefname{inequality}{inequality}{inequalities}

\usepackage{ifthen}
\usepackage{microtype}

\providecommand{\texorpdfstring}[2]{#1}

\newcommand{\dss}{distributed storage system\xspace}
\newcommand{\chunks}{chunks\xspace}
\newcommand{\node}{node\xspace}
\newcommand{\nodes}{nodes\xspace}
\newcommand{\stripe}{stripe\xspace}
\newcommand{\stripes}{stripes\xspace}

\newcommand{\etal}{et al.\xspace}
\newcommand{\ie}{i.e.,\xspace}
\newcommand{\eg}{e.g.,\xspace}

\newcommand{\codename}{convertible code\xspace}
\newcommand{\codenames}{\codename{}s\xspace}

\newcommand{\convertible}[2]{({#1},{#2})\text{-optimally convertible}}

\newcommand{\Code}{\mathcal{C}}
\newcommand{\Cs}{\varsigma}

\newcommand{\IdMat}{\mathbf{I}}
\newcommand{\V}[1]{\mathbf{#1}}

\newcommand{\fieldq}{\mathbb{F}_{q}}
\newcommand{\NN}{\mathbb{N}^+}

\newcommand{\ICode}{\Code^{\Initial}}
\newcommand{\FCode}{\Code^{\Final}}

\newcommand{\Gen}[1]{\mathbf{G}^{#1}}
\newcommand{\IGen}{\Gen{\Initial}}

\newcommand{\Par}[1]{\mathbf{P}^{#1}}
\newcommand{\IPar}{\Par{\Initial}}

\newcommand{\Col}[2]{\V{g}^{#1}_{#2}}

\newcommand{\Colp}[3]{\V{\tilde{g}}^{#1}_{{#2},{#3}}}
\newcommand{\IColp}[2]{\Colp{\Initial}{#1}{#2}}

\newcommand{\Down}{\mathcal{D}}
\newcommand{\Vectors}{\mathcal{V}}
\newcommand{\Vectorss}{\mathcal{W}}

\newcommand{\StripeCols}[2]{\mathcal{S}^{#1}_{#2}}

\newcommand{\StripeICols}[1]{\StripeCols{\Initial}{#1}}
\newcommand{\StripeFCols}[1]{\StripeCols{\Final}{#1}}

\newcommand{\AllCols}[1]{\mathcal{S}^{#1}}
\newcommand{\AllICols}{\AllCols{\Initial}}
\newcommand{\AllFCols}{\AllCols{\Final}}
\newcommand{\NewCols}[1]{\mathcal{N}_{#1}}
\newcommand{\AllNewCols}{\mathcal{N}}

\newcommand{\Access}{\mathcal{A}}
\newcommand{\StripeAccess}[1]{\Access_{#1}}

\newcommand{\UnchangedCols}{\mathcal{U}}
\newcommand{\UnchangedStripesCols}[2]{\mathcal{U}_{{#1},{#2}}}

\newcommand{\ParamFormat}[4]{({#1},\allowbreak{#2};\allowbreak{#3},\allowbreak{#4})}

\newcommand{\ParamsDefault}{\ParamFormat{\In}{\Ik}{\Fn}{\Fk}}

\newcommand{\CodeDefault}{\(\ParamsDefault\) \codename}
\newcommand{\RegimeCodeDefault}{\(\ParamFormat{\In}{\Ik = \Cs \Fk}{\Fn}{\Fk}\) \codename}
\newcommand{\RegimeCodesDefault}{\RegimeCodeDefault{}s\xspace}

\newcommand{\mergecode}{\(\ParamFormat{\In}{\Ik}{\Fn}{\Fk = \Cs\Ik}\) \codename}
\newcommand{\splitcode}{\(\ParamFormat{\In}{\Ik = \Cs\Fk}{\Fn}{\Fk}\) \codename}
\newcommand{\genmergecode}{\(\ParamFormat{\In}{\{\Ikg{i}\}_{i=1}^{\Is}}{\Fn}{\Fk = \sum_{i=1}^{\Is} \Ikg{i}}\) convertible code\xspace}

\newcommand{\gensplitcode}{\(\ParamFormat{\In}{\Ik = \sum_{i=1}^{\Fs} \Fkg{i}}{\Fn}{\{\Fkg{i}\}_{i=1}^{\Fs}}\) convertible code\xspace}
\newcommand{\gensplitcodes}{\gensplitcode{}s\xspace}

\newcommand{\Partition}{\mathcal{P}}
\newcommand{\IPart}{\Partition^{\Initial}}
\newcommand{\FPart}{\Partition^{\Final}}

\newcommand{\K}[1]{k^{#1}}
\newcommand{\N}[1]{n^{#1}}
\newcommand{\R}[1]{r^{#1}}
\newcommand{\Ss}[1]{\lambda^{#1}}
\newcommand{\C}[1]{\Code^{#1}}
\newcommand{\Set}[2]{{P^{#1}_{#2}}}

\newcommand{\Initial}{I}
\newcommand{\Ik}{{\K{\Initial}}}
\newcommand{\In}{{\N{\Initial}}}
\newcommand{\Ir}{{\R{\Initial}}}
\newcommand{\Is}{{\Ss{\Initial}}}
\newcommand{\IC}{{\C{\Initial}}}
\newcommand{\initialcode}{[\In, \Ik]}
\newcommand{\splitinitialcode}{[\In, \Ik = \Cs\Fk]}
\newcommand{\Iset}[1]{{\Set{\Initial}{#1}}}
\newcommand{\Ikg}[1]{{k^{\Initial}_{#1}}}
\newcommand{\Ikmax}{{k^{\Initial}_{*}}}
\newcommand{\Ikgs}{{k^{\Initial}_{s}}}
\newcommand{\Ings}{{n^{\Initial}_{s}}}
\newcommand{\Cgs}{{\C s}}

\newcommand{\Final}{F}
\newcommand{\Fk}{{\K{\Final}}}
\newcommand{\Fn}{{\N{\Final}}}
\newcommand{\Fr}{{\R{\Final}}}
\newcommand{\Fs}{{\Ss{\Final}}}
\newcommand{\Fsd}{{\hat{\lambda}^{\Final}}}
\newcommand{\FC}{{\C{\Final}}}
\newcommand{\finalcode}{[\Fn, \Fk]}
\newcommand{\Fset}[1]{{\Set{\Final}{#1}}}
\newcommand{\Fkg}[1]{{k^{\Final}_{#1}}}
\newcommand{\Fkmax}{{k^{\Final}_{*}}}
\newcommand{\Fkgm}{{k^{\Final}_{m}}}
\newcommand{\Fngm}{{n^{\Final}_{m}}}

\newcommand{\Mk}[2]{{k_{{#1},{#2}}}}
\newcommand{\Mkmax}[1]{{k_{{#1},*}}}

\newcommand{\size}{M}
\newcommand{\msg}{\V{m}}

\newcommand{\IT}{\lozenge}

\newcommand{\floorfrac}[2]{\left\lfloor\sfrac{#1}{#2}\right\rfloor}
\newcommand{\ceilfrac}[2]{\left\lceil\sfrac{#1}{#2}\right\rceil}
\newcommand{\Mod}[2]{{#1}\bmod{#2}}
\DeclareMathOperator{\lcm}{lcm}
\DeclareMathOperator{\rank}{rk}
\DeclareMathOperator{\Span}{span}
\DeclareMathOperator{\argmax}{argmax}

\usepackage{amsthm}

\newtheorem{theorem}{Theorem}
\newtheorem{lemma}[theorem]{Lemma}
\newtheorem{prop}[theorem]{Proposition}

\theoremstyle{definition}
\newtheorem{definition}{Definition}

\theoremstyle{remark}

\crefname{lemma}{Lemma}{Lemmas}
\crefname{prop}{Proposition}{Propositions}
\crefname{corollary}{Corollary}{Corollaries}
\crefname{definition}{Definition}{Definitions}
\crefname{example}{Example}{Examples}
\crefname{remark}{Remark}{Remarks}

\ProvideDocumentEnvironment{proofenv}{o}
{\IfNoValueTF{#1}%
    {\begin{proof}}%
    {\begin{proof}[#1]}%
}
{\end{proof}}
\usepackage{fancyhdr}
\fancypagestyle{firststyle}
{
   \fancyhf{}
   
   \chead{\footnotesize This is an extended version of an IEEE ISIT 2020 paper with the same title.}
}

\begin{document}

\title{Access-optimal Linear MDS Convertible Codes\\ for All Parameters\\
\thanks{
This work was funded in part by an NSF CAREER award (CAREER-1943409) and a Google faculty research award. We thank Michael Rudow for his suggestions in the writing of this paper.
}
}

\author{Francisco~Maturana, V.~S.~Chaitanya~Mukka, and~K.~V.~Rashmi%
\thanks{F. Maturana and K. V. Rashmi are with the Computer Science Department at Carnegie Mellon University, Pittsburgh, PA, USA.
(emails: fmaturan@cs.cmu.edu and rvinayak@cs.cmu.edu).}%
\thanks{V. S. Chaitanya Mukka is with the Department of CS \& IS at BITS Pilani, Goa Campus, Goa, India. (email: f20150536@goa.bits-pilani.ac.in)}
}

\maketitle

\thispagestyle{firststyle}

\begin{abstract}
In large-scale distributed storage systems, erasure codes are used to achieve fault tolerance in the face of node failures. Tuning code parameters to observed failure rates has been shown to significantly reduce storage cost. Such tuning of redundancy requires \textit{code conversion}, i.e., a change in code dimension and length on already encoded data. \textit{Convertible codes}~\cite{maturana20convertible} are a new class of codes designed to perform such conversions efficiently. The \textit{access cost} of conversion is the number of nodes accessed during conversion.

Existing literature has characterized the {access cost} of conversion of linear MDS convertible codes only for a specific and small subset of parameters.
In this paper, we present lower bounds on the access cost of conversion of linear MDS codes for \textit{all} valid parameters. Furthermore, we show that these lower bounds are tight by presenting an explicit construction for access-optimal linear MDS convertible codes for all valid parameters.
En route, we show that, one of the degrees-of-freedom in the design of convertible codes that was inconsequential in the previously studied parameter regimes, turns out to be crucial when going beyond these regimes and adds to the challenge in the analysis and code construction.
\end{abstract}

\section{Introduction}\label{sec:introduction}

Erasure codes are an essential tool for providing resilience against \node failures in a \dss~\cite{ghemawat2003google,facebookECsavings2010_forACM,huang2012erasure,hadoophdfsec,rashmi2013hotstorage,asterisxoring,rashmi2014hitchhiker}.
When using an $[n, k]$ erasure code, $k$ \chunks of data are encoded into $n$ \chunks, called a \textit{\stripe}.
These chunks are then distributed among $n$ different ``\nodes'' in the system, where nodes correspond to distinct storage devices typically residing on distinct servers.
For the purposes of theoretical study, each stripe can be viewed as a \textit{codeword}, by viewing each of the $n$ chunks as one of the $n$ codeword symbols.
The parameters $n$ and $k$ are usually chosen based on node failure rate, which might vary over time.
Redundancy tuning, i.e., changing $n$ and $k$ in response to fluctuations in the failure rate of storage devices can achieve significant savings (11\% to 44\%) in storage space~\cite{kadekodi2019cluster}.
Due to practical system constraints, changing $n$ alone is typically insufficient and both $n$ and $k$ have to be changed simultaneously~\cite{kadekodi2019cluster}.
The resource cost of changing $n$ and $k$ on already encoded data can be prohibitively high and is a key barrier in the practical adoption of redundancy tuning~\cite{maturana20convertible}.
Other reasons to change $n$ and $k$ on already encoded data might include variations in data popularity, failure rate uncertainty, or restrictions on the total amount of used storage.

The \emph{code conversion} problem defined in~\cite{maturana20convertible} involves converting multiple stripes of an $\initialcode$ code (denoted by $\IC$) into (potentially multiple) stripes of an $\finalcode$ code (denoted by $\FC$), along with desired constraints on decodability such as both codes being Maximum Distance Separable (MDS).
Considering multiple stripes enables code conversions to allow for changes in the code dimension (from $\Ik$ to $\Fk$).
\textit{Convertible codes}~\cite{maturana20convertible} are code pairs that enable code conversion, usually designed to minimize the cost of conversion.
A detailed description of the convertible codes framework is provided in \cref{sec:convertible}.

There are several ways in which one might measure the cost of conversion.
We focus on the \textit{access cost} of conversion, which is measured in terms of the total number of nodes that need to be accessed during conversion.
In~\cite{maturana20convertible}, the authors focus on the so-called \textit{merge} regime, wherein multiple initial stripes are merged into one.
Specifically, they consider the case where $\Fk = \Cs\Ik$ for some integer $\Cs \geq 2$, and propose explicit constructions for converible codes that achieve optimal access cost for the merge regime. 
We review these results for the merge regime in \cref{sec:merge}.

The results presented in this work are two fold. (1) We present lower bounds on the access cost of conversion for linear MDS codes \textit{for all valid parameters}, {that is, all $\In, \Ik, \Fn, \Fk \in \NN$ such that $\In > \Ik$ and $\Fn > \Fk$}. (2) We show that the proposed lower bounds are tight by presenting an \textit{explicit construction} of linear MDS convertible codes that is access optimal for all parameter regimes.
To achieve this, we first define and study the \textit{split regime} in \cref{sec:split}, where $\Ik = \Cs\Fk$ for an integer $\Cs \geq 2$, that is, a single initial stripe is split into multiple final stripes.
We prove a (tight) lower bound on the access cost of conversion in the split regime, and describe a conversion procedure which has optimal access cost when used with any systematic MDS code.
We then present in \cref{sec:general} a tight lower bound on the access cost of conversion for linear MDS convertible codes for all valid parameters (termed \textit{general regime}) by reducing conversion in the general regime to a combination of generalizations of conversions in the split and merge regimes. While the split and the merge regimes might seem somewhat restrictive, we show that, perhaps surprisingly, the proposed conversion procedure for the general regime that builds on top of the generalized split and merge regime is \textit{optimal}.
Interestingly, one of the degrees-of-freedom in the design of convertible codes (called ``partitions'' described subsequently in \cref{sec:convertible}), which is inconsequential in the split and merge regimes, turns out to be crucial in the general regime. The proposed construction for access-optimal convertible codes for the general regime builds on the constructions for split and merge regimes, while separately optimizing along this additional degree-of-freedom.

\section{Background and Related work}\label{sec:background}

\subsection{Convertible codes \texorpdfstring{~\cite{maturana20convertible}}{}}\label{sec:convertible}

A \emph{conversion} from an $\initialcode$ initial code $\IC$ to an $\finalcode$ final code $\FC$ is a procedure that takes as input a set of initial \stripes from $\IC$ and outputs a set of final \stripes from $\FC$, such that the final \stripes together encode the same information as the initial \stripes.
To avoid degeneracy, $\Fn > \Fk$ and $\In > \Ik$ is assumed.
Let $\fieldq$ be a finite field, and consider a message $\msg \in \fieldq^\size$, where $\size = \lcm(\Ik, \Fk)$.
The number of initial stripes is $\Is = \sfrac{\size}{\Ik}$ and the number of final stripes is $\Fs = \sfrac{\size}{\Fk}$.
Let $[n] = \{1, \ldots, n\}$, $\Ir = \In - \Ik$ and $\Fr = \Fn - \Fk$.
Let $\msg[S]$ denote the projection of $\msg$ onto the coordinates in the set $S$, and let $\Code(\msg)$ denote the encoding of $\msg$ under code $\Code$.
Consider an \textit{initial partition} $\IPart = \{\Iset{1}, \ldots, \Iset{\Is}\}$ of $[\size]$ such that $|\Iset{i}| = \Ik\ (\forall i \in [\Is])$, and a \textit{final partition} $\FPart = \{\Fset{1}, \ldots, \Fset{\Fs}\}$ of $[\size]$ such that $|\Fset{j}| = \Fk\ (\forall j \in [\Fs])$.
These partitions determine how message symbols are mapped to each of the initial and final stripes.
For example, the $i$-th initial stripe will only encode the symbols of $\msg$ indexed by $\Iset{i}$.

\begin{definition}[Convertible code~\cite{maturana20convertible}]
    An \CodeDefault over $\fieldq$ is defined by:
    \begin{enumerate}
        \item 
        a pair of codes $(\IC, \FC)$ over $\fieldq$ such that $\IC$ is $\initialcode$ and $\FC$ is $\finalcode$;
        \item
        a pair of partitions $(\IPart, \FPart)$ of $[\size = \lcm(\Ik,\Fk)]$ such that $|\Iset{i}| = \Ik$ for all $\Iset{i} \in \IPart$ and $|\Fset{j}| = \Fk$ for all $\Fset{j} \in \FPart$; and
        \item
        a conversion procedure which, for any $\msg \in \fieldq^{\size}$, takes the set of initial codewords $\{\IC(\msg[\Iset{i}]) \mid \Iset{i} \in \IPart\}$ as input, and outputs the corresponding set of final codewords $\{\FC(\msg[\Fset{j}]) \mid \Fset{j} \in \FPart\}$.
    \end{enumerate}
\end{definition}
In this paper, we will restrict our focus to the case where $\IC$ and $\FC$ are both linear and MDS.

The \emph{access cost} of a conversion procedure is the total number of \nodes that are read or written during conversion.
Recall that each node in a stripe corresponds to a single symbol from the corresponding codeword, therefore access cost is equivalent to the number of codeword symbols that are read or written during conversion.
We distinguish three types of \nodes during conversion: \emph{unchanged \nodes}, which remain \emph{as is} during the conversion process, and are present in both the initial and final configuration (possibly in different stripes); \emph{retired \nodes}, which are present in the initial configuration and throughout the conversion, but not in the final configuration; and \emph{new \nodes}, which are introduced during conversion, and are present in the final configuration, but not in the initial configuration.
Unchanged and retired \nodes may be accessed for reading during conversion, and new \nodes are always accessed for writing during the conversion.
A \codename that maximizes the number of unchanged \nodes is said to be \emph{stable}.

The \emph{read access set} of an \CodeDefault is a set of tuples $\Down \subseteq [\Is] \times [\In]$, where $(i, j) \in \Down$ corresponds to the $j$-th \node in initial \stripe $i$.
After a conversion, each new \node holds a fixed linear combination of the contents of the \nodes indexed by $\Down$.
We denote the accessed nodes from initial stripe $i$ as $\Down_i = \{j \mid (i,j) \in \Down\}$.
Thus, the access cost of a conversion with read access set of size $d = |\Down|$ and $m$ new \nodes is $d + m$.
Clearly, there always exists a conversion procedure with read access cost $\size$, which reconstructs the original message $\msg$ and re-encodes according to $\FCode$.
We refer to this procedure as the \emph{default approach}.

An \CodeDefault is \emph{access-optimal} if and only if it achieves the minimum access cost over all \CodeDefault{}s.

\subsection{Merge regime\texorpdfstring{~\cite{maturana20convertible}}{}}\label{sec:merge}

The \emph{merge regime} is the subset of valid parameter values for \codenames where $\Fk = \Cs\Ik$, for some integer $\Cs \geq 2$.
Thus, in this regime we have $\Is = \Cs$ and $\Fs = 1$.
This regime was the focus of \cite{maturana20convertible}, wherein the following lower bound on access cost was shown.
\begin{theorem}[\hspace{1sp}\cite{maturana20convertible}]
    For all linear MDS \CodeDefault, the access cost of conversion is at least $\Fr + \Cs\min\{\Ik,\Fr\}$.
    Furthermore, if $\Ir < \Fr$, the access cost of conversion is at least $\Fr + \Cs\Ik$.
\end{theorem}
An explicit construction for access-optimal \codenames for all values in the merge regime was also provided in \cite{maturana20convertible}.

\subsection{Other related works}
 
The closest related work~\cite{maturana20convertible} proposes the convertible codes framework considered in this work (discussed at length above).
Several other works in the literature~\cite{rashmi2011enabling,rai2015adaptive,hu2018generalized,sonowal2017adaptive,MZT18} have considered variants of the code conversion problem, largely within the context of so-called ``regenerating codes''~\cite{DGWWR10}. The study on regenerating codes, which are a class of codes that optimize for recovery for a small subset of nodes within a stripe (as opposed to decoding all original data), was initiated by Dimakis~\etal~\cite{DGWWR10}. Subsequently numerous works have studied and constructed optimal regenerating codes (\eg~\cite{papailiopoulos2013repairTransactions,cadambe2011polynomial,tamo2013zigzag,guruswami2016repairing,rashmi2011optimal,sasidharan2015high,shah2012distributed,ye2017explicit,mahdaviani2018product,shah2011interference,suh2011journal,rouayheb2010fractional,chowdhury2018newconstructions,dau2017optimal,rashmi2017piggybacking}).
Specific instances of code conversion can be viewed as instances of the repair problem, for example, increasing $n$ while keeping $k$ fixed as studied in~\cite{rashmi2011enabling,MZT18}.

In a recent work~\cite{su2020local}, Su et al.\ study a related problem in the context of coded computation for distributed matrix multiplication.
In~\cite{su2020local}, Su et al.\ propose a coding scheme for coded matrix-multiplication with the property that certain changes to the code parameters only require local re-encoding of the data stored in each server.

\subsection{Notation}

This subsection introduces notation that generalizes the notation used in \cite{maturana20convertible} and is used throughout this paper.
Let $\Gen{\IT} = (\Col{\IT}{1} \cdots \Col{\IT}{\N{\IT}}) \in \fieldq^{\K{\IT} \times \N{\IT}}$ be a generator matrix of MDS code $\C{\IT}$ for $\IT \in \{\Initial, \Final\}$.
An encoding vector in relation to $\msg \in \fieldq^{\size}$ is associated to each node in the initial or final stripes.
The encoding vector $\Colp{\IT}{i}{j} \in \fieldq^{\size}$ of node $j \in [\N{\IT}]$ in stripe $i \in [\Ss{\IT}]$ with partition set $\Set{\IT}{i} \in \Partition^{\IT}$ is defined such that $\Colp{\IT}{i}{j}[\Set{\IT}{i}] = \Col{\IT}{j}$, and 0 everywhere outside of $\Set{\IT}{i}$.
The difference between $\Col{\IT}{j}$ and $\Colp{\IT}{i}{j}$ is that the former describes the encoding of the $j$-th symbol relative to the information encoded in a single initial (resp.\ final) stripe, while the latter describes the encoding of the $j$-th symbol of the $i$-th initial (resp.\ final) stripe relative to the information jointly encoded by all initial (resp.\ final) stripes (i.e., the message $\msg$).

Let $\StripeCols{\IT}{i} = \{\Colp{\IT}{i}{j} \mid j \in [\N{\IT}]\}$ be the encoding vectors for a particular stripe, and let $\AllCols{\IT} = \bigcup_{i \in [\Ss{\IT}]} \StripeCols{\IT}{i}$.
Let $\UnchangedCols = \AllICols \cap \AllFCols$ be the encoding vectors of unchanged nodes, and define $\UnchangedStripesCols{i}{j} = \StripeICols{i} \cap \StripeFCols{j}$, where the index $i$ or $j$ is dropped if $\Is = 1$ or $\Fs = 1$, respectively.
Let $\StripeAccess{i} = \{\IColp{i}{j} \mid j \in \Down_i\}$ be the encoding vectors of nodes that are read from initial stripe $i$, and define $\Access = \{\IColp{i}{j} \mid (i,j) \in \Down\}$ as the set of all encoding vectors of nodes that are read.
Finally, let $\AllNewCols = \AllFCols \setminus \AllICols$ be the encoding vectors of new nodes, and define $\NewCols{i} = \StripeFCols{i} \setminus \AllICols$ as the encoding vectors of new nodes of a particular stripe $i$.
Notice that it must hold that $\AllNewCols \subseteq \Span(\Access)$.
For simplicity, we sometimes refer to a node and its encoding vector interchangeably.

\section{Split regime}\label{sec:split}

The \emph{split regime} of \codenames corresponds to the case where a single initial \stripe is split into multiple final \stripes.
{This regime is, in some sense, the opposite of the merge regime, in which multiple initial \stripes are combined into one final \stripe.}
Specifically, an \CodeDefault is in the split regime if $\Ik = \Cs\Fk$ for an integer $\Cs \geq 2$, with arbitrary $\In$ and $\Fn$.
Notice that in this regime we have that $\size = \lcm(\Ik,\Fk) = \Ik$ and thus $\Is = 1$ and $\Fs = \Cs$.

First, in \cref{sec:split-lower-bound}, we show a lower bound on access cost for the split regime.
In \cref{sec:split-upper-bound} we show a matching upper bound on access cost by showing that for every systematic $[\In, \Cs\Fk]$ MDS code $\Code$ there exists an access-optimal \splitcode having $\Code$ as its initial code by presenting a conversion procedure whose cost matches the lower bound.

\subsection{Access cost lower bound for the split regime}\label{sec:split-lower-bound}

In this subsection, we lower bound the access cost of conversion in the split regime. 
This is done by first showing a lower bound on write access cost, and then showing a lower bound on the read access cost of conversion.

The following fact simplifies the analysis of the split regime.
\begin{prop}
For a linear MDS \RegimeCodeDefault, all possible pairs of initial and final partitions are equivalent (up to relabeling). 
\end{prop}
\begin{proofenv}
There is only one possible initial partition $\IPart = \{[\Ik]\}$, hence any two final partitions can be made equivalent by relabeling \nodes.
\end{proofenv}
Therefore, we do not need to consider differences in partitions in our analysis of the split regime.

\begin{prop}\label[prop]{prop:split_stable}
In a linear MDS \RegimeCodeDefault, there are at most $\Fk$ unchanged nodes in each of the final stripes (i.e., at least $\Fr$ new nodes per stripe).
Hence, there are at most $\Ik$ unchanged nodes in total. 
\end{prop}

\begin{proofenv}
For any final stripe $i \in [\Cs]$, any subset $\mathcal{V} \subseteq \StripeFCols{i}$ of size at least $\Fk + 1$ is linearly dependent due to the MDS property.
Thus, $\mathcal{V} \subseteq \AllICols$ contradicts the fact that $\ICode$ is MDS.
Hence, each final stripe $i$ has at most $\Fk$ unchanged nodes.
\end{proofenv}

Therefore, the total write access cost in the split regime is at least $\Cs\Fr$.

Now we focus on bounding the read access cost.
The general strategy we use to obtain bounds on read access cost is to consider a specially chosen set $\Vectorss$ of $\Fk$ nodes from a final stripe, which by the MDS property of the final code is enough to decode all data in that stripe.
We then use the fact that final stripes are the result of conversion to identify a set $\Vectors$ of initial nodes that contain all the information contained in $\Vectorss$.
The MDS property of the initial code constrains the information available in $\Vectors$, which allows us to derive a lower bound on its size and thus a lower bound on the number of read nodes.

\begin{lemma}\label[lemma]{lem:split_less_par}
For all linear MDS \RegimeCodesDefault, the read access set $\Down$ satisfies $|\Down| \geq (\Cs-1)\Fk + \min\{\Fr, \Fk\}$. 
\end{lemma}

\begin{proofenv}
If $\Fr \geq \Fk$, then all data should be decodable by accessing only new nodes in the final stripes, and the result follows easily since all data must have been read to create the new nodes.
Therefore, assume for the rest of this proof that $\Fr < \Fk$.

Suppose, for the sake of contradiction, that $|\Down| < (\Cs-1)\Fk + \Fr$.
Let $u$ be a node in some final stripe $i \in [\Cs]$ which is neither read nor written.
Such a stripe and node exist since otherwise every node in the final stripes would be accessed (for either read or write) and thus $\AllFCols$ would be in the span of $\Access$, which is a contradiction since $\rank(\AllFCols) = \Ik$.

Let $\Vectorss_1$ be a subset of nodes of the same final stripe $i$ such that $\Vectorss_1 \subseteq \NewCols{i}$ and $|\Vectorss_1| = \Fr$.
Such a subset exists by virtue of \cref{prop:split_stable}.
Further, let $\Vectorss_2 \subseteq \StripeFCols{i} \setminus (\Vectorss_1 \cup \{u\})$ be such that $|\Vectorss_2| = \Fk - \Fr$. 
Clearly $\Vectorss = \Vectorss_1 \cup \Vectorss_2$ is of size $|\Vectorss| = \Fk$ and can reconstruct the contents of $u$, by the MDS property of the final code.
In other words, $u \in \Span (\Vectorss)$.

Let $\Vectorss'_2 = (\Vectorss_2 \cap \UnchangedCols_i)$ be the unchanged nodes in $\Vectorss_2$.
Since $\Vectorss_1$ and $\Vectorss_2 \setminus \Vectorss'_2$ only have new nodes, they are both contained in $\Span(\Access)$, therefore $\Vectorss \subseteq \Span (\Access \cup \Vectorss'_2)$.
Notice that the subset $\Vectors = (\Access \cup \Vectorss'_2)$ consists only of initial nodes.
Furthermore, it holds that $\rank(\Access)\leq |\Down|$ and $\rank(\Vectorss'_2) \leq |\Vectorss_2| = \Fk - \Fr < \Fk$.
Thus:
\[
    \rank(\Vectors) \leq \rank (\Access) + \rank (\Vectorss'_2) \leq |\Down| + (\Fk - \Fr) < \Ik.
\]
This implies that $\Vectorss$ is spanned by less than $\Ik$ initial nodes (which do not include $u$).
However, by the MDS property of the initial code, any subset of less than $\Ik$ initial nodes that does not contain node $u$, has no information about $u$.
This causes a contradiction with the fact that $u \in \Span(\Vectorss) \subseteq \Span(\Vectors)$.
Thus, we must have $|\Down| \geq (\Cs-1)\Fk + \Fr$.
\end{proofenv}

It is easy to show that if we only read unchanged nodes, it is not possible to do better than the default approach.
This follows from the fact that unchanged nodes are already present in the final stripes and hence using them to create the new nodes will contradict with the MDS property.
Retired nodes, on the other hand, do not have this drawback.
Thus, intuitively, based on \cref{lem:split_less_par}, one might expect to achieve an efficient conversion by reading from the retired nodes.
However, we next show that it is not possible to achieve lower read access cost than the default approach when $\Ir < \Fr$.

\begin{lemma}\label[lemma]{lem:split_more_par}
    For all linear MDS  $(n^I, k^I = \Cs k^F; n^F, k^F )$ convertible codes, if $r^I  < r^F$ then the read access set $\mathcal{D}$ satisfies $|\mathcal{D}| \geq \Cs k^F$.
\end{lemma}

\begin{proofenv}
Suppose, for the sake of contradiction, that $|\Down| < \Cs\Fk$.
Let $u$ be a node in some final stripe $i \in [\Cs]$ which is neither read nor written.
Such a stripe and node always exist as described in the proof of \cref{lem:split_less_par}.
We will choose a subset of nodes $\Vectorss \subseteq \StripeFCols{i}$ of size $|\Vectorss| = \Fk$.
By the MDS property of the final code, node $u$ is decodable from $\Vectorss$, \ie $u \in \Span (\Vectorss)$.
There are two cases for the choice of $\Vectorss$ depending on the total number of accessed nodes in stripe $i$:

\noindent\textbf{Case 1:} If $|\NewCols{i}| +|\UnchangedCols_i \cap \Access| \geq \Fk$, then let $\Vectorss \subseteq \NewCols{i} \cup (\UnchangedCols_i \cap \Access)$.
That is, $\Vectorss$ only contains nodes that are read or written.
It is easy to see that $\Vectorss \subseteq \Span(\Access)$. 

Clearly, $\Access$ contains only initial nodes, and the following holds:
\[
    \rank (\Access) \leq |\Down| < \Cs \Fk = \Ik.
\]
However, this is a contradiction with the fact that $u \in \Span(\Vectorss)$, since by the MDS property of the initial code, $\Access$ contains no information about node $u$.

\noindent\textbf{Case 2:} If $|\NewCols{i}| + |\UnchangedCols_i \cap \Access| < \Fk$, then choose $\Vectorss = (\Vectorss_1 \cup \Vectorss_2)$, where $\Vectorss_1 = (\NewCols{i} \cup (\UnchangedCols_i \cap \Access))$ and $\Vectorss_2$ is any subset of $(\StripeFCols{i} \setminus (\Vectorss_1 \cup \{u\}))$ of size $|\Vectorss_2| = \Fk - |\Vectorss_1|$.
That is, $\Vectorss$ contains all the nodes of final stripe $i$ that are read or written (in addition to other unchanged nodes distinct from $u$).
It is easy to see that $\Vectorss_1 \subseteq \Span(\Access)$ and thus $\Vectorss \subseteq \Span(\Access \cup \Vectorss_2)$.
Furthermore, the subset $\Vectors = (\Access \cup \Vectorss_2)$ consists only of initial nodes.

Notice that there are at most $(|\AllICols| - |\UnchangedCols_i|) = (\Ik + \Ir - |\UnchangedCols_i|)$ read nodes outside of final stripe $i$ (i.e., in $\AllICols \setminus \UnchangedCols_i$).
Therefore, we can bound $\rank(\Access)$ by $\rank(\Access) \leq \Ik + \Ir - |\UnchangedCols_i| + |\UnchangedCols_i \cap \Access|$.
On the other hand, it is clear that $\rank(\Vectorss_2) \leq |\Vectorss_2| = \Fk - |\NewCols{i}| - |\UnchangedCols_i \cap \Access|$.
Combining these, we get:
\begin{align*}
    \rank(\Vectors) &\leq \rank(\Access) + \rank(\Vectorss_2) \\
    &\leq \Ik + \Ir + \Fk - |\UnchangedCols_i| - |\NewCols{i}| \\
    &\leq \Ik + \Ir - \Fr \\
    &< \Ik.
\end{align*}
However, this is a contradiction with the fact that $u \in \Span(\Vectorss) \subseteq \Span(\Vectors)$, since by the MDS property of the initial codes, $\Vectors$ contains no information about node $u$.
\end{proofenv}

By combining all the results in this subsection, we obtain the following lower bound on the access cost of conversion in the split regime.

\begin{theorem} \label{thm:split-lower-bound}
    The total access cost of any linear MDS \RegimeCodeDefault is at least $(\Cs - 1)\Fk + \min\{\Fr, \Fk\} + \Cs\Fr$ if $\Ir \geq \Fr$, and at least $\Cs\Fn$ otherwise.
\end{theorem}
\begin{proofenv}
Follows from \cref{prop:split_stable,lem:split_less_par,lem:split_more_par}.
\end{proofenv}

As we show in the next subsection, this lower bound is tight since it is achievable.

\subsection{Access-optimal \codenames for the split regime}\label{sec:split-upper-bound}

In this subsection we present a construction of access-optimal convertible codes in the split regime.
Under this construction, any systematic MDS code can be used as the initial code. The final code corresponds to the projection of the initial code onto the coordinates of any $\Fk$ systematic nodes. 
Since our construction can be applied to existing codes and only specifies the conversion procedure, we introduce the following definition capturing the property of codes that can be converted efficiently.
\begin{definition}
A code $\IC$ is \emph{$\convertible{\Fn}{\Fk}$} if and only if there exists an $[\Fn, \Fk]$ code $\FC$ (along with partitions and conversion procedure) that form an access-optimal \CodeDefault.
\end{definition}
{The conversion procedure that leads to optimal access cost (meeting the lower bound in \cref{thm:split-lower-bound}) is as follows.}

\textit{Conversion procedure:} All the systematic nodes are used as unchanged nodes. When $\Ir < \Fr$ or $\Fr \geq \Fk$, the conversion is trivial since one cannot do better than the default approach. 
The conversion procedure for the nontrivial case proceeds as follows.
For all but one final stripe, all unchanged nodes are read ($(\Cs - 1)\Fk$ in total), and the new nodes are naively constructed from them.
For the remaining final stripe, $\Fr$ retired nodes are read, and then the unchanged nodes from the other final stripes are used to remove their interference from the retired nodes to obtain $\Fr$ new nodes.

\begin{theorem}\label{thm:split-upper-bound}
    Every systematic linear MDS $\splitinitialcode$ code $\IC$ is $\convertible{\Fn}{\Fk}$.
\end{theorem}
\begin{proofenv}
    If $\Fr > \min\{\Ir, \Fk\}$, then the default approach achieves the bound stated in \cref{thm:split-lower-bound}.
    Thus, assume $\Fr \leq \min\{\Ir, \Fk\}$.
    Let $\IGen = [\IdMat \mid \IPar]$ be the generator matrix of $\IC$ and assume nodes are numbered in the same order as the columns of $\IGen$.
    Define $\FC$ as the code generated by the matrix formed by taking the first $\Fk$ rows of $\IGen$, and columns $1, \ldots, \Fk$ and $\Ik + 1, \ldots, \Ik + \Fr$.
    Let $(i - 1)\Fk + 1, \ldots, i\Fk$ be the columns of the unchanged nodes corresponding to final stripe $i \in [\Cs]$.
    Consider the following conversion procedure: read the the subset of unchanged nodes $U = \{\Fk + 1, \ldots, \Cs\Fk\}$ and the retired nodes $R = \{\Ik + 1, \ldots, \Ik + \Fr\}$.
    To construct the new nodes for stripe 1, simply project the nodes of $R$ onto their first $\Fk$ coordinates by using nodes $U$.
    To construct the new nodes for stripe $i \neq 1$, simply use then nodes in $U$.
    This conversion procedure reads a total of $|U| + |R| = (\Cs - 1)\Fk + \Fr$ \nodes and writes a total of $\Cs\Fr$ new nodes, which matches the bound from \cref{thm:split-lower-bound}.
\end{proofenv}

Notice that convertible codes created using the construction above are stable.
We show this property is, in fact, necessary.
\begin{lemma}
    All access-optimal convertible codes for the split regime are stable.
\end{lemma}
\begin{proofenv}
    \Cref{thm:split-upper-bound} shows that there exist stable access-optimal codes for the split regime.
    Since any unstable convertible code must incur higher write access cost and at least as much read access cost, it cannot be access-optimal.
\end{proofenv}

\section{General regime}\label{sec:general}

In this section, we will study the general regime of \codenames with arbitrary valid parameter values (i.e.\ any $\In > \Ik$ and $\Fn > \Fk$).
Recall that the choice of partition functions was inconsequential in the split and merge regimes. In contrast, it turns out that \textit{the choice of initial and final partitions play an important role in the general regime}. This makes the general regime significantly harder to analyze.
We deal with this complexity by reducing conversion in the general regime to generalized versions of the split and merge conversions, and by \textit{identifying the conditions on initial and final partitions to minimize total access cost}.

In \cref{sec:general-splitmerge}, we explore a generalization of the split regime and of the merge regime.
In \cref{sec:general-lower}, these generalizations are used to lower bound the access cost of conversion in the general regime.
In \cref{sec:general-upper}, we describe a conversion procedure and construction for access-optimal conversion in the general regime which utilizes ideas from the constructions for generalizations of split and merge regimes.

\subsection{Generalized split and merge regimes} \label{sec:general-splitmerge}
The generalized split and merge regimes are similar to the split and merge regimes, except that the generalized variants allow for initial or final stripes of unequal sizes.
This flexibility enables the generalized split and merge regimes to be used as building blocks in the analysis of the general regime.
In these generalized variants, the message length $\size$ is defined to be $\max\{\Ik, \Fk\}$ (which coincides with the definition of $\size$ in the split and merge regime), but now the sets in the initial and final partitions need not be all of the same size.

Since the initial (or final) stripes might be of different lengths, we define them as shortenings of a common code $\Code$.
\begin{definition}
An $s$-shortening of an $[n,k]$ code $\Code$ is the code $\Code'$ formed by all the codewords in $\Code$ that have 0 in a fixed subset of $s$ positions, with those $s$ positions deleted.
\end{definition}
Shortening a code has the effect of decreasing the length $n$ and dimension $k$ while keeping $(n - k)$ fixed.
It can be shown that an $s$-shortening of an $[n, k]$ MDS code is an $[n - s, k - s]$ MDS code.
\emph{Lengthening} is the inverse operation of shortening, and has the effect of increasing length $n$ and dimension $k$ while keeping $(n - k)$ fixed.
For linear codes, an $s$-lengthening of a code can be defined as adding $s$ additional columns to its parity check matrix.
Similarly, it can be shown that for an $[n, k]$ MDS code, there exists an $s$-lengthening of it that is an $[n + s, k + s]$ MDS code (assuming a large enough field size).

\subsubsection{Generalized split regime} \label{sec:generalized-split}

In the generalized split regime, $\Is = 1$ is fixed, $\Fs > 1$ is arbitrary, and the final partition $\FPart = \{\Fset{1}, \ldots, \Fset{\Fs}\}$ is such that $|\Fset{i}| = \Fkg{i}$ and $\sum_{i \in [\Fs]} \Fkg{i} = \Ik$.
Let $\Fkmax = \max_{i \in [\Fs]} \Fkg{i}$.
Then $\FC$ is a $[\Fn,\Fkmax]$ MDS code, and the code corresponding to each final stripe is some fixed shortening of $\FC$.
In this case, we define $\Fr = \Fn - \Fkmax$.

\begin{definition}
A \gensplitcode for the generalized split regime is a variant of a convertible code defined by:
\begin{enumerate}
    \item
    $\IC$ and $\FC$ as $\initialcode$ and $[\Fn,\Fkmax]$ codes, where $\Fkmax = \max_{i \in [\Fs]} \Fkg{i}$,
    \item
    a partition $\FPart = \{\Fset{1}, \ldots, \Fset{\Fs}\}$ where $|\Fset{i}|=\Fkg{i}$, 
and 
    \item
    a conversion procedure such that each final stripe $i$, is an $s_i$-shortening of $\FC$ where $s_i=\Fkmax - \Fkg{i}$.
\end{enumerate}
\end{definition}

The generalized split regime has an access cost lower bound similar to the split regime presented in \cref{sec:split}.
We show this by showing that a more efficient conversion procedure for the generalized split regime would imply the existence of a conversion procedure for split regime violating \cref{thm:split-lower-bound}.

\begin{theorem}\label{thm:gen-split-lower-bound}
  For all linear MDS \gensplitcodes, the read access set $\Down$ satisfies:
  \[
  |\Down| \geq \Ik - \max \{\Fkmax - \Fr, 0\},
  \text{ where } \Fkmax = \max_{i \in [\Fs]} \Fkg{i}.
  \]
\end{theorem}

\begin{proofenv}
Suppose, for the sake of contradiction, that there exists a conversion procedure with read access cost $|\Down| < \Ik - \max \{\Fkmax - \Fr, 0\}$ for some convertible code in the generalized split regime with codes $\ICode$ and $\FCode$.
We modify the initial code $\IC$ by lengthening it to an $[\Ings, \Ikgs]$ MDS code $\Cgs$, such that $\Ikgs = \Fs \Fkmax$ and $\Ir = \In -\Ik = \Ings - \Ikgs$.
This adds $\sum_{i=1}^\Fs (\Fkmax - \Fkg{i}) = (\Ikgs - \Ik)$ extra ``pseudo-nodes'' to the initial code, which we denote with $\Vectorss$.

We then define a new conversion procedure from code $\Cgs$ to code $\FCode$ which uses the conversion procedure for the generalized split regime convertible code as a subroutine, and then simply reads all the added pseudo-nodes to construct the new nodes.
This procedure only reads the read access set $\Down$ from $\Cgs$ along with the $(\Ikgs - \Ik)$ pseudo-nodes.

Hence, the total read access is,
\begin{align*}
    |\Down \cup \Vectorss| &< (\Ik - \max \{\Fkmax - \Fr, 0\}) + (\Ikgs - \Ik) \\
    &\leq (\Fs -1)\Fkmax + \min\{\Fr, \Fkmax\}.
\end{align*}
However, the codes $\Cgs$ and $\FCode$ with the new conversion procedure clearly form an MDS $(\Ings, \Ikgs = \Fs \Fkmax; \Fn, \Fkmax)$ convertible code.
Therefore, this is in contradiction to \cref{thm:split-lower-bound}.
Then, it must hold that $|\Down| \geq \Ik - \max \{\Fkmax - \Fr, 0\}$.
\end{proofenv}

This lower bound is achievable for all pairs of initial and final parameters.
Similar to the case of the split regime, shown in \cref{sec:split-upper-bound}, we can use any systematic MDS codes as initial and final codes, and access all but a set of nodes of size $\Fkmax$ (forming the largest final stripe) to perform this conversion, as described below.

\textit{Conversion procedure:}
All the systematic nodes are used as unchanged nodes.
When $\Ir < \Fr$ or $\Fr \geq \Fkmax$, the conversion is trivial since one cannot do better than the default approach.
The conversion procedure for the nontrivial case proceeds as follows.
For all but the largest final stripe, all unchanged nodes are read ($\Cs\Fk - \Fkmax$ in total), and the new nodes are naively constructed from them.
For the largest final stripe, the $\Fr$ retired nodes are read, and then the unchanged nodes from the other final stripes are used to remove their interference from the retired nodes to obtain $\Fr$ new nodes.

\subsubsection{Generalized merge regime} \label{sec:generalized-merge}
In the generalized merge regime, the sets in the initial partition need not be all of the same size.
In this case, we fix $\size = \Fk$ and $\Fs = 1$, while $\Is > 1$ is arbitrary.
The initial partition $\IPart = \{\Iset{1}, \ldots, \Iset{\Is}\}$ is such that $|\Iset{i}| = \Ikg{i}$ and $\sum_{i \in [\Is]} \Ikg{i} = \Fk$.
Let $\Ikmax = \max_{i \in [\Is]} \Ikg{i}$.
Then $\IC$ is a $[\In,\Ikmax]$ MDS code, $\Ir = \In - \Ikmax$, and the code corresponding to each initial stripe is some fixed shortening of $\IC$.

\begin{definition}
    A \genmergecode for the generalized merge regime is a variant of a convertible code defined by: 
    \begin{enumerate}
        \item $\IC, \FC$ as $[\In,\Ikmax]$ and $\finalcode$ codes, where $\Ikmax = \max_{i \in [\Is]} \Ikg{i}$ 
        \item partition $\IPart = \{\Iset{1}, \ldots, \Iset{\Is}\}$ where $|\Iset{i}|=\Ikg{i}$, and
        \item a conversion procedure such that each initial stripe $i$, is an $s_i$-shortening of $\IC$ where $s_i=\Ikmax - \Ikg{i}$.
    \end{enumerate}
\end{definition}

The next theorem gives a lower bound on the read access cost of a \genmergecode.

\begin{theorem}\label{thm:gen-merge-lower-bound}
    For all \genmergecode, $|\Down_i| \geq \min\{\Ikg{i}, \Fr\}$ for all $i \in [\Is]$.
    Furthermore, if $\Ir < \Fr$, then $|\Down_i| \geq \Ikg{i}$ for all $i \in [\Is]$.
\end{theorem}
\begin{proofenv}
    Follows from the proofs of Lemmas 10, 11, and 13 in \cite{maturana20convertible}, with some straightforward modifications to account for the difference in the number of nodes of each initial stripe.
\end{proofenv}

We can achieve this lower bound by shortening an access-optimal $\ParamFormat{\In}{\Ikmax}{\Fngm}{\Fkgm}$ \codename, where $\Fkgm = \Is\Ikmax$ and $\Fngm = \Fkgm + \Fr$.

\subsection{Access cost lower bound for the general regime} \label{sec:general-lower}

In this subsection, we study the access cost lower bound for conversions in the general regime (i.e., for all valid parameter values, $\In > \Ik$ and $\Fn > \Fk$).
As in the merge and split regime, we show that when $\Ir \geq \Fr$, significant reduction in access cost can be achieved.
However when $\Ir < \Fr$, one cannot do better than the default approach.

For an \CodeDefault with $\Ik \neq \Fk$ and partitions $(\IPart, \FPart)$, let $\Mk{i}{j} = |\Iset{i} \cap \Fset{j}|$ for $(i, j) \in [\Is] \times [\Fs]$ and let $\Mkmax{i} = \max_{j \in [\Fs]} \Mk{i}{j}$.
\begin{lemma} \label[lemma]{thm:general-stripe-lower-bound}
    For all linear MDS \CodeDefault{s} with $\Ik \neq \Fk$:
    \[
      |\Down_i| \geq \Ik  - \max\{\Mkmax{i} - \Fr, 0\}\text{ for all }i \in [\Is].
    \]
    Moreover, if $\Ir < \Fr$ then $|\Down_i| \geq \Ik$ for all $i \in [\Is]$.
\end{lemma}

    \begin{proofenv}
    Let $i \in [\Is]$ be an initial stripe.
    There are two cases.
    
    \noindent\textbf{Case $\Mkmax{i} < \Ik$:}
    In this case, we can reduce this conversion to a conversion in the generalized split regime by focusing on initial stripe $i$, and considering messages which are zero everywhere outside of $\Iset{i}$.
    This is equivalent to a $\ParamFormat{\In}{\Ik}{\Mkmax{i} + \Fr}{\{\Mk{i}{j}\}_{j=1}^{\Fs}}$ convertible code.
    Then, the result follows from \cref{thm:gen-split-lower-bound}.
    
    \noindent\textbf{Case $\Mkmax{i} = \Ik$:}
    Let $j = \argmax_{j' \in [\Fs]} \Mk{i}{j'}$.
    In this case, we can reduce this conversion to conversion in the generalized merge regime by focusing on final stripe $j$, and considering messages which are zero everywhere outside of $\Fset{j}$.
    This is equivalent to a $\ParamFormat{\In}{\{\Mk{i}{j}\}_{i=1}^{\Is}}{\Fn}{\Fk}$ convertible code.
    Then, the result follows from \cref{thm:gen-merge-lower-bound}.
    \end{proofenv}

We prove a lower bound on the total access cost of conversion in the general regime by using \cref{thm:general-stripe-lower-bound} on all initial stripes and finding a partition that minimizes the value of the sum.

\begin{theorem} \label{thm:general-lower-bound}
    For every linear MDS \CodeDefault such that $\Ik \neq \Fk$, it holds that: \[
        |\Down| \geq \Is\Fr + (\Mod{\Is}{\Fs})(\Ik - \max\{\Mod{\Fk}{\Ik}, \Fr\})
    \]
    if $\Fr < \min\{\Ik, \Fk\}$.
    Furthermore, if $\Ir < \Fr$ or $\Fr \geq \min\{\Ik, \Fk\}$, then $|\Down| \geq \size$.
\end{theorem}

\begin{proofenv}
    Clearly, it holds that $|\Down| = \sum_{i=1}^{\Is} |\Down_i|$.
    Then, the case $\Ir < \Fr$ follows directly from \cref{thm:general-stripe-lower-bound}.
    Otherwise, by the same lemma we have:
    \begin{equation}\label[inequality]{eq:general-lower-bound}
        |\Down| = \sum_{i=1}^{\Is} |\Down_i| \geq \sum_{i=1}^{\Is} \Ik - \max \{\Mkmax{i} - \Fr, 0\}.
    \end{equation}
    First, we consider the case $\Ik > \Fk$.
    Notice that in this case $(\Mod{\Is}{\Fs}) = \Is$ and $(\Mod{\Fk}{\Ik}) = \Fk$.
    If $\Fr \geq \Fk$, then the result is trivial, so assume $\Fr < \Fk$.
    Since $\Mkmax{i} \leq \Fk$ for all $i \in [\Is]$, we have:
    \[
        |\Down| \geq \sum_{i=1}^{\Is} \Ik - \max \{\Mkmax{i} - \Fr, 0\} \geq \Is (\Ik + \Fr - \Fk),
    \]
    which proves the result.
    
    Now, we consider the case $\Ik < \Fk$.
    Assume, for now, that the right hand side of \cref{eq:general-lower-bound} is minimized when:
    \begin{equation}\label{eq:minimizer}
        \Mkmax{i} =
        \begin{cases}
        \Ik, &\text{ for } 1 \leq i \leq (\Is - (\Mod{\Is}{\Fs})) \\
        (\Mod{\Fk}{\Ik}), &\text{ otherwise.}
        \end{cases}
    \end{equation}
    Then, from \cref{eq:general-lower-bound} we have:
    \begin{equation} \label[inequality]{eq:general-merge-case}
        |\Down| \geq \Is\Ik - (\Is - (\Mod{\Is}{\Fs}))\max \{\Ik - \Fr, 0\} - (\Mod{\Is}{\Fs})\max \{(\Mod{\Fk}{\Ik}) - \Fr, 0\}
    \end{equation}
    If $\Fr \geq \Ik$, then the result is trivial, so assume $\Fr < \Ik$.
    Then, by manipulating the terms of \cref{eq:general-merge-case}, the result is obtained.
    
    It only remains to prove that the right hand side of \cref{eq:general-lower-bound} is minimized when \cref{eq:minimizer} holds.
    
    Notice that this is equivalent to showing that $s = \sum_{i=1}^{\Is} \max\{\Mkmax{i} - \Fr, 0\}$ is maximized by the proposed assignment.
    To prove this, we will show that any optimal assignment to the variables $\Mk{i}{j}$ can be modified to be of the proposed form, without decreasing the value of the objective $s$.
    Firstly, it is straightforward to check that there exists a feasible assignment to the variables $\Mk{i}{j}$ that satisfies the statement.
    
    Suppose we have an optimal assignment for variable $\Mk{i}{j}$ that is not of the proposed form and assume, without loss of generality, that $\Mkmax{1} \geq \cdots \geq \Mkmax{\Is}$.
    Let $1 \leq i \leq (\Is - (\Mod{\Is}{\Fs}))$ be the least such that $\Mkmax{i} < \Ik$, and let $j = \argmax_{j' \in [\Fs]} \Mk{i}{j'}$.
    It must hold that $\Mkmax{i} > \max\{\Fr, \Mod{\Fk}{\Ik}\}$, otherwise this assignment could not be optimal.
    Notice that $\Mkmax{i'} = \Ik$ for all $i' < i$ and since $\Ik \nmid (\Fk - \Mkmax{i})$, there exists at least one $i' > i$ such that $\Mk{i'}{j} > 0$.
    Furthermore, there exists $j' \neq j$ such that $\Mk{i}{j'} > 0$, since $\Mkmax{i} < \Ik$.
    Then, we can ``swap'' elements from $\Mk{i}{j'}$ with $\Mk{i'}{j}$.
    This increases $\Mkmax{i}$ and decreases $\Mkmax{i'}$ by at most the same amount.
    Since $\Mkmax{i} > \Fr$, this cannot decrease the value of the objective $s$.
    We can repeat this procedure until $\Mkmax{i} = \Ik$ for all $1 \leq i \leq (\Is - (\Mod{\Is}{\Fs}))$.
    
    Notice now that for every $(\Is - (\Mod{\Is}{\Fs})) \leq i \leq \Is$ it holds that:
    \begin{equation}\label[inequality]{eq:general-helper}
        \Mkmax{i} \leq \Mod{\Fk}{\Ik}
    \end{equation} otherwise, there must exist some $j \in [\Fs]$ such that $\sum_{i=1}^{\Is} \Mk{i}{j} > \Fk$.
    If $\Fr < (\Mod{\Fk}{\Ik})$, then \cref{eq:general-helper} must hold with equality.
    Otherwise, each such $\Mkmax{i}$ will contribute exactly $\Fr$ to the objective $s$, so they can be modified to be of the desired form without decreasing $s$.
\end{proofenv}

\subsection{Access-optimal convertible codes for the general regime} \label{sec:general-upper}
In this subsection we prove that the lower bound from \cref{thm:general-lower-bound} is achievable by presenting convertible code constructions that are access-optimal in the general regime.
We first present the conversion procedure for our construction and then describe the construction of the initial and final codes that are compatible with this conversion procedure.

\begin{figure}
    \centering
    \includegraphics[width=.8\textwidth]{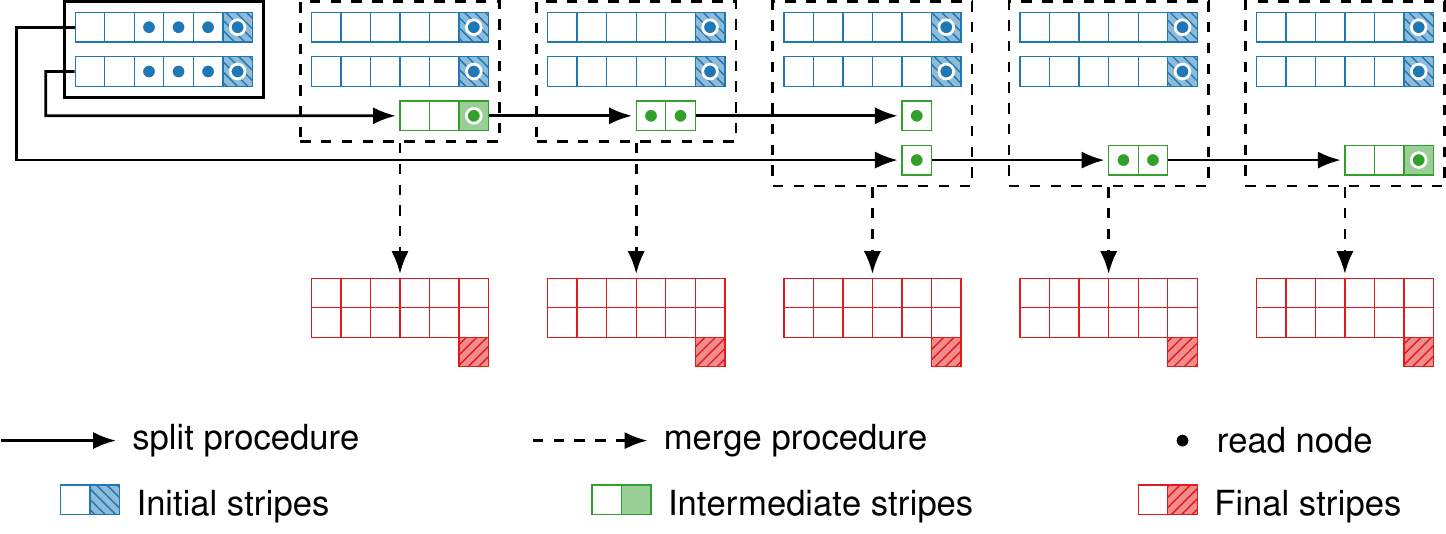}
    \caption{
        Conversion procedure from $[6,5]$ to $[13,12]$ {($\Is = 12$ and $\Fs = 5$)}.
        Read access cost is $18$ compared to $60$ in the default approach ($70\%$ savings).
    }
    \label{fig:general-merge}
\end{figure}

\begin{figure}
    \centering
    \includegraphics[width=.8\textwidth]{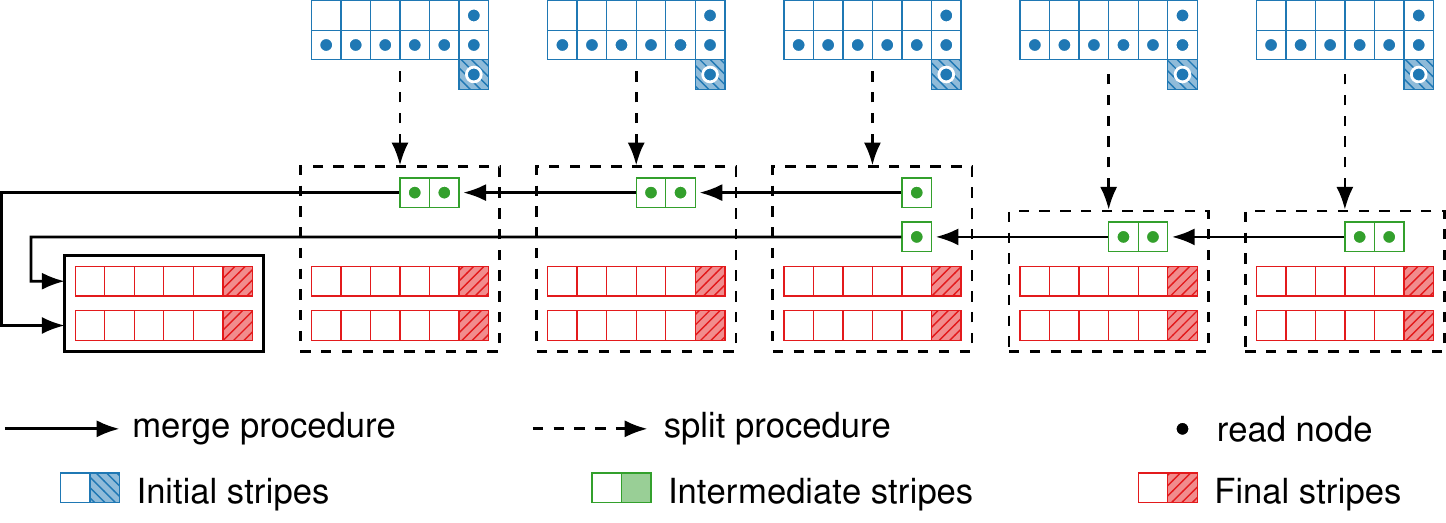}%
    \caption{
        \hspace{-4.75pt} Conversion procedure from $[13,12]$ to $[6,5]$ {($\Is = 5$ and $\Fs = 12$)}.
        Read access cost is $40$, compared to $60$ in the default approach ($33.\overline{3}\%$ savings in read access cost).
    }
    \label{fig:general-split}
\end{figure}

\subsubsection{Conversion procedure} \label{sec:general-conversion}

Conversion in the general regime can be achieved by combining the conversion procedures of codes in the generalized split and merge regimes.
In the case where $\Ir < \Fr$, we access $\Ik$ nodes from each initial stripe and use the default approach.
For the case where $\Ir \geq \Fr$, we present the conversion procedure by considering three cases: $\Ik = \Fk$, $\Ik < \Fk$, and $\Ik > \Fk$.

\noindent\textbf{Case $\Ik = \Fk$:}
Notice that $\In \geq \Fn$ since $\Ir \geq \Fr$.
This is a degenerate case where any $\Fn$ \nodes from the initial stripe can be kept unchanged.

\noindent\textbf{Case $\Ik < \Fk$:}
We will separate the nodes of initial stripes into $\Fs$ disjoint groups with the same amount of information.
This requires splitting some initial stripes into what we call \emph{intermediate stripes}, which are then assigned to different groups.
We will finally merge each group to form the $\Fs$ final stripes.
Specifically (see \cref{fig:general-merge}):
\begin{enumerate}
    \item Assign $\floorfrac{\Fk}{\Ik}$ initial stripes to each group (dashed boxes in \cref{fig:general-merge}).
    \item \label{step:split}
    Use an \(\ParamFormat{\In}{\Ik}{\Fn}{\{\Fkg{i}\}_{i=1}^{\Fsd}}\) conversion procedure to (generalized) split the $(\Mod{\Is}{\Fs})$ remaining initial stripes to obtain $\Fsd$ intermediate stripes, where $\Fsd = \ceilfrac{\Ik}{(\Mod{\Fk}{\Ik})}$, $\Fkg{i} = (\Mod{\Fk}{\Ik}$) for $i \in [\Fsd - 1]$, and $\Fkg{\Fsd} = (\Mod{\Fk}{\Ik})$ if $(\Mod{\Fk}{\Ik}) \mid \Ik$ and $\Fkg{\Fsd} = (\Mod{\Ik}{(\Mod{\Fk}{\Ik})})$ otherwise.
    Each intermediate stripe is assigned to a different group.
    \item \label{step:merge}
    The conversion procedure for generalized merge is used to turn each stripe group into a single final stripe.
\end{enumerate}

The total number of nodes read during conversion is $\Is \Fr + (\Mod{\Is}{\Fs})(\Ik - \max\{\Mod{\Fk}{\Ik}, \Fr\})$, which matches \cref{thm:general-lower-bound}.

\noindent\textbf{Case $\Ik > \Fk$:}
Conversion occurs in two steps (see \cref{fig:general-split}):
\begin{enumerate}
    \item {
    First, use an \(\ParamFormat{\In}{\Ik}{\Fn}{\{\Fkg{i}\}_{i=1}^{\Fsd}}\) conversion procedure to (generalized) split each initial stripe, where $\Fsd = (\ceilfrac{\Ik}{\Fk})$, $\Fkg{i} = \Fk$ for $i \in [\Fsd - 1]$ (corresponding to final stripes), and $\Fkg{\Fsd} = \Fk$ if $\Fk \mid \Ik$ (corresponding to another final stripe) and $\Fkg{\Fsd} = (\Mod{\Fk}{\Ik})$ otherwise (corresponding to an intermediate stripe).
    }
    
    \item Assemble the $\Is (\Mod{\Fk}{\Ik})$ remaining nodes from the intermediate stripes into $(\Mod{\Fs}{\Is})$ final stripes.
    This is done using the default approach, since all the remaining nodes would have been already accessed in the first step.
\end{enumerate}

The total number of nodes read in this case during conversion is $\Is (\Fr + \Ik - \Fk)$, which matches \cref{thm:general-lower-bound}.

Therefore, the total access cost of conversion when $\Ir \geq \Fr$ and $\Ik \neq \Fk$ is $(\Is + \Fs) \Fr + (\Mod{\Is}{\Fs})(\Ik - \max\{\Mod{\Fk}{\Ik}, \Fr\})$, while the access cost of the default approach is $\Fs\Fn$.

\subsubsection{Access-optimal construction} \label{sec:general-construction}

Since the conversion procedure in \cref{sec:general-conversion} is based on the generalized split and merge regimes, we only need to ensure that the constructed codes can perform those conversions with optimal access cost.

\begin{theorem}
    For all $\Fk \leq \Ik$, every systematic linear MDS $\initialcode$ code $\IC$ is $\convertible{\Fn}{\Fk}$. For all $\Fk \leq \Cs\Ik$ with integer $\Cs > 2$, every access-optimal systematic linear MDS \mergecode is $\convertible{\Fn}{\Fk}$.
\end{theorem}

\begin{proofenv}
    Recall, from \cref{sec:generalized-split} that any systematic $\initialcode$ code $\IC$ can be used as the initial code of an access-optimal convertible code in the generalized split regime (i.e., an \gensplitcode).
    Since the conversion procedure for the general regime in the case where $\Ik > \Fk$ only uses conversions from the generalized split regime and conversions from the generalized merge regime that can be carried out using the default approach, it is clear that any systematic code $\IC$ can be used.
    Similarly, from \cref{sec:generalized-merge} we know that any $\initialcode$ code $\IC$ that is $\convertible{\Fn}{\Cs\Ik}$ for an integer $\Cs \geq 2$ can achieve conversion with optimal access cost in a \genmergecode, where $\Is \leq \Cs$.
    Since the conversion procedure for the general regime in the case where $\Ik < \Fk$ only uses conversions from the generalized split and merge regimes, it is clear that any $\convertible{\Fn}{\Cs\Ik}$ code $\IC$ such that $\Cs \geq \lceil \sfrac{\Fk}{\Ik} \rceil$ can be used.
\end{proofenv}

Therefore, the constructions for the merge regime presented in \cite{maturana20convertible} can be used to construct access-optimal \codenames in the general regime.

\bibliographystyle{IEEEtran}
\bibliography{IEEEabrv,main}

\end{document}